\documentclass[nohyper,12pt,letterpaper]{JHEP3}
\usepackage{amsmath}
\date{\today}

\title{Nonlocal Effects on D-branes in Plane-Wave Backgrounds}

\author{Ori J. Ganor and Uday Varadarajan\\ \\
Department of Physics \\ 366 LeConte Hall \\
University of California \\ Berkeley, CA 94720 \\ \\
and\\ \\
Lawrence Berkeley National Labs\\
Berkeley, CA 94720 \\ \\
Emails: \email{origa,udayv@socrates.berkeley.edu}}

\abstract{

We argue that the effective field theory on D3-branes in a plane-wave
background with 3-form flux is a nonlocal deformation of Yang-Mills
theory. In the case of NSNS flux, it is a dipole field theory with
lightlike dipole vectors. For an RR 3-form flux the dipole theory is
strongly coupled. We propose a weakly coupled S-dual description for it.
The S-dual description is local at any finite order in string
perturbation theory but becomes nonlocal when all perturbation theory
orders are summed together.

}

\keywords{String theory, AdS/CFT, pp-waves, nonlocality, dipole theory}

\received{October 2, 2002}

\preprint{\hepth{0210035}\\ UCB-PTH-02-40\\ LBNL-51535}
\begin{document}

\def\be{\begin{equation}} 
\def\ee{\end{equation}} 
\def\bear{\begin{eqnarray}} 
\def\eear{\end{eqnarray}} 
\def\nn{\nonumber} 
 
\newcommand\bra[1]{{\langle {#1}|}} 
\newcommand\ket[1]{{|{#1}\rangle}} 
 
\def\tr{{\mbox{tr}}} 
\def\Atr{{\mbox{Tr}}} 

\def\a{\alpha} 
\def\b{\beta} 
\def\g{\gamma} 
\def\d{\delta} 
\def\lam{\lambda} 
\def\u{\mu} 
\def\v{\nu} 
\def\r{\rho} 
\def\t{\tau}
\def\z{\zeta}
\def\s{\sigma}
\def\th{\theta} 

\def\Qh{\hat{Q}}
\def\baret{\overline{\eta}}
\def\homega{{\hat{\omega}}}
\def\bpsi{{\overline{\psi}}}
\def\wtau{{\widetilde{\tau}}} 
\def\bth{{\overline{\theta}}} 
\def\blam{{\overline{\lambda}}} 

\def\da{{\dot{\a}}}
\def\db{{\dot{\b}}}
\def\dg{{\dot{\g}}}
 
\def\bj{{\overline{j}}} 
\def\bk{{\overline{k}}} 
\def\bz{{\overline{z}}} 
\def\sa{{\hat{a}}}
\def\sb{{\hat{b}}}
\def\sc{{\hat{c}}}
\def\wa{{\tilde{a}}}
\def\wb{{\tilde{b}}}
\def\wc{{\tilde{c}}}
\def\oa{{\overline{a}}}
\def\ob{{\overline{b}}}
\def\oc{{\overline{c}}}
\def\od{{\overline{d}}}

\def\wdg{{\wedge}} 

\newcommand\ev[1]{{\langle {#1}\rangle}}
\newcommand\SUSY[1]{{{\cal N} = {#1}}}
\newcommand\diag[1]{{\mbox{diag}({#1})}}
\newcommand\com[2]{{\left\lbrack {#1}, {#2}\right\rbrack}}

\newcommand\px[1]{{\partial_{#1}}} 
\newcommand\qx[1]{{\partial^{#1}}} 

\newcommand\rep[1]{{\bf {#1}}}

\def\gam{{\widetilde{\gamma}}} 
\def\sig{{\sigma}} 
\def\hsig{{\hat{\sigma}}} 

\def\eps{{\epsilon}} 

\def\bZ{{\overline{Z}}}
\def\BR{{\mathbb R}}
\def\Lag{{\cal L}}
\def\cO{{\cal O}}
\def\wcO{{\widetilde{\cal O}}}
\def\vL{{\vec{L}}}
\def\vx{{\vec{x}}}
\def\vy{{\vec{y}}}

\def\vLf{{\vec{\lambda}}}

\newcommand\cvL[1]{{L^{#1}}} 

\def\npsi{{\psi^{(inv)}}}
\def\bnpsi{{\bpsi^{(inv)}}}
\def\wpsi{{\widetilde{\psi}}}
\def\bwpsi{{\overline{\widetilde{\psi}}}}
\def\hpsi{{\hat{\psi}}}
\def\bhpsi{{\overline{\hat{\psi}}}}

\newcommand\nDF[1]{{{D^F}_{#1}}}
\def\wA{{\widetilde{A}}}
\def\wF{{\widetilde{F}}}
\def\wJ{{\widetilde{J}}}
\def\hA{{\hat{A}}}
\def\hF{{\hat{F}}}
\def\hJ{{\hat{J}}}
\def\MapL{{\Upsilon}}
\def\hR{{\hat{R}}}

\def\cpl{{\lambda}}  
\def\mcr{{\mathcal{R}}}
\def\xv{{\vec{x}}}
\def\xvt{{\vec{x}^{\top}}}
\def\nv{{\hat{n}}}
\def\nvt{{\hat{n}^{\top}}}
\def\hM{{M}}
\def\hgM{{\widetilde{M}}}
\def\utr{{\mbox{tr}}}

\def\vwsX{{\bf X}} 
\def\dvwsX{{\dot{\bf X}}} 
\newcommand\vxmod[1]{{{\bf X}_{#1}}} 
\newcommand\dvxmod[1]{{\dot{{\bf X}}_{#1}}} 

\def\vwsP{{\bf \Pi}} 

\def\wsX{{X}} 
\def\wsZ{{Z}} 
\def\bwsZ{{\overline{Z}}} 
\def\wsPsi{{\Psi}} 
\def\bwsPsi{{\overline{\Psi}}} 

\def\twsZ{{\widetilde{Z}}} 
\def\twS{{\widetilde{S}}} 
\def\btwsZ{{\overline{\widetilde{Z}}}} 
\def\twsPsi{{\widetilde{\Psi}}} 
\def\btwsPsi{{\overline{\widetilde{\Psi}}}} 
\def\tV{{\widetilde{V}}}
\def\talp{{{\tilde{\a}}'}} 

\def\hLf{{\hat{\lambda}}}
\def\hL{{\hat{L}}}


\section{Introduction}
The restrictions imposed by the conditions of Lorentz invariance and
locality play central roles in our understanding of the formal
properties of quantum field theories.  However, in string theory
neither of these conditions appears to be fundamental. Thus, it is
interesting to consider simple situations where they are relaxed. In
particular, we will examine the properties of D-branes in certain
plane-wave backgrounds with strong 3-form fields. As we will show in
detail, the low energy effective theory describing the fluctuations of
these D-branes is a non-local, Lorentz violating dipole theory
\cite{Bergman:2000cw}-\cite{Bergman:2001rw}.

Typical interaction terms in the Lagrangian of this field theory
are of the form $\int \phi_1(\vx)\phi_2(\vx+\vL_1)\phi_3(x+\vL_1 +
\vL_2)\cdots d^4\vx$ where $\phi_i$ are fields and the $\vL_i$ are
fixed world-volume vectors. Roughly speaking, the non-locally coupled fields
$\phi_i$ correspond to stretched open strings with end-points that are
separated by $\vL_i$ and with angular momentum along planes
transverse to the brane.  These strings are stabilized by the presence
of strong 3-form fluxes with legs aligned along the dipole vectors as
well as the plane of rotation \cite{Dasgupta:2000ry}.

An exciting application of string theory with strong 3-form field
strengths is the $AdS_3/CFT_2$ correspondence \cite{Maldacena:1997re}.
Unfortunately, progress had been limited by the fact that string
theory in $AdS$ backgrounds with RR field strengths
are difficult to analyze exactly. However, the authors of
\cite{Berenstein:2002jq} have shown that a particularly tractable
limit of the $AdS/CFT$ correspondence can be obtained by taking the
Penrose limit of type-IIB string theory on $AdS_5\times S^5$ to obtain
a plane-wave background. They were able to precisely match the
properties of a certain subsector of $\SUSY{4}$ Super-Yang-Mills CFT
(operators with large R-charge) with the exact results of
\cite{Metsaev:2001bj}-\cite{Russo:2002rq} for strings in plane-wave
backgrounds.

Similarly, one can consider the Penrose limits of $AdS_3\times
S^3\times T^4$ \cite{Lunin:2002fw,Gomis:2002qi}. As IIB has two
three-form field strengths, $H^1$ (NSNS) and $H^2$ (RR), one finds a
pair of models which are related by S-duality. The Penrose limit of
the theory with $H^1$ flux is
\bear
ds^2 & =& dx^{+} dx^{-} + \mu x^i x^i (dx^+)^2 
        - dx^a dx^a - dx^i dx^i,
\label{ppNSi}\\
H^{1} & =& -\mu dx^{+}\wdg (dx^6\wdg dx^7 + dx^8 \wdg dx^9), 
\label{ppNSii}\\
e^{\phi} & =& g_s,
 \label{ppNSiii}
\eear
where $ds^2$ is the interval in string frame,
$x^{\pm} = x^0 \pm x^1$, the $x^a$
are coordinates on $T^4$ with $a=2,\ldots,5$ and $i=6,\ldots,9$.
The Penrose limit of the S-dual configuration is
\bear
ds^2 & =& dx^{+} dx^{-} + \mu x^i x^i (dx^{+})^2 - dx^a dx^a - dx^i dx^i,
\label{ppRi}\\
H^{2} & =& \mu dx^{+}\wdg (dx^6\wdg dx^7 + dx^8 \wdg dx^9),
\label{ppRii}\\
e^{\phi} & =& \frac{1}{g_s}.
\label{ppRiii}
\eear
Exact results for the spectrum of both models were obtained in
\cite{Lunin:2002fw,Gomis:2002qi}.
Further, open strings and
D-branes in these and other plane-wave backgrounds have been studied
in \cite{Billo:2002ff}-\cite{Hyun:2002fk}.


In this paper we will study the interactions of the low energy
effective theory of the D-brane excitations. We will show that $N$
D3-brane probes of the plane-wave background 
(\ref{ppNSi})-(\ref{ppNSiii}) are exactly
described at low energies by a nonlocal $U(N)$ dipole gauge theory
\cite{Bergman:2000cw} with a lightlike dipole vector $\vL$
proportional to $\mu$.

A more complicated problem is the description of $N$ D3-brane probes
of the pp-wave background (\ref{ppRi})-(\ref{ppRiii}),
which has RR flux. It is
related to the S-dual description of the lightlike dipole theory.  We
attack this problem by first studying the S-dual description of a
$U(1)$ lightlike dipole theory and then guessing the generalization of
that result to a $U(N)$ gauge group.  We find that in any finite order
of string perturbation theory the interactions of the D3-brane probes
of the pp-wave background (\ref{ppRi})-(\ref{ppRiii}) are local.
Yet our result
suggests that summing the local interactions to all orders in
perturbation theory exhibits an intrinsic nonlocality
with a characteristic length proportional to the string coupling
constant, $g_s$.

The paper is organized as follows.
In section \ref{secDT} we review the definition and salient features
of dipole theories.
In section \ref{secLLppNS} we identify the lightlike dipole
theory as the low energy description of D3-branes in 
the pp-wave background (\ref{ppNSi})-(\ref{ppNSiii}).
In section \ref{secPropS} we analyze the S-dual of the $U(1)$
lightlike dipole theories and conjecture an extension of the result
to $U(N)$,
We conclude in section \ref{secDisc} with a list of possible
extensions of our work.

\section{Definition of dipole theories and their salient features}
\label{secDT}
The dipole field theories that we will work with in this paper are
nonlocal field theories that are deformations of $\SUSY{4}$ SYM.
The Lagrangian of $\SUSY{4}$ SYM is
\bear
\Lag_{\SUSY{4}} &=& \frac{1}{g^2}{\mbox{tr}}\left\{
\frac{1}{4}F_{\u\v}F^{\u\v}
+\frac{1}{2}\sum_{I=1}^6 D_\u\Phi^I D^\u\Phi^I
+i\sum_{a=1}^4 \bpsi^\da_a {\sigma^{\u\a}}_\da D_\u\psi_\a^a
\right\}
\nn\\ &+&\frac{1}{g^2}{\mbox{tr}}\left\{
\sum_{I<J} \com{\Phi^I}{\Phi^J}^2
+\epsilon^{\a\b}\sum_{I,a,b}\gamma^I_{ab}\Phi^I\psi_\a^a\psi_\b^b
+\epsilon_{\da\db}\sum_{I,a,b}
\gamma^{I ab}\Phi^I\bpsi^\da_a\bpsi^\db_b
\right\},
\nn\\
D_\u\Phi^I &\equiv& \px{\u}\Phi^I + i\com{A_\u}{\Phi^I}.
\label{LagSYM}
\eear
Here $\Phi^I$ ($I=1\dots 6$) are adjoint scalar fields of $U(N)$ which
transform as a vector of the R-symmetry group $Spin(6)$.  The
$\psi_\a^a$ ($a=1\dots 4$) are adjoint Weyl fermions in the ${\bf 4}$
of $Spin(6)$. Their complex conjugate fields $\bpsi^\da_a$ transform
in the complex conjugate representation ${\bf \overline{4}}$ of
$Spin(6)$. $\gamma^{I ab}$ are the Clebsch-Gordan coefficients of
$Spin(6)$ and ${\sigma^{\u\a}}_\da$ are Pauli matrices.

The dipole theories are obtained from $\SUSY{4}$ SYM by the following steps
(see \cite{Bergman:2001rw} for more details):
\begin{enumerate}
\item
Define the complex linear combinations of the 6 scalar fields
of (\ref{LagSYM}):
$$
Z_k\equiv \Phi_{2k-1} + i\Phi_{2k},\qquad
\bZ_k\equiv \Phi_{2k-1} -i\Phi_{2k},\qquad k=1,2,3,
$$
and assign a constant space-time
4-vector $\vL_k$ to each scalar field $Z_k$.

\item
Modify the covariant derivatives of the scalar fields so that
$D_\u Z_k$ at the space-time point $x$ will be:
\be\label{defDu}
D_\u Z_k(x)\equiv \px{\u}Z_k(x) 
-i A_\u(x-\frac{1}{2}\vL_k) Z_k(x)
+i Z_k(x) A_\u(x+\frac{1}{2}\vL_k).
\ee
Note that the fields $Z_k$ are $N\times N$ matrices in the adjoint
representation of $U(N)$.  Thus, equation (\ref{defDu}) implies that
the quanta of the fields $Z_k$ are dipoles whose ends are at
$x\pm\frac{1}{2}\vL_k$.  The gauge transformation of the scalar fields
is
$$
Z_k(x)\mapsto
\Omega^{-1}(x-\frac{1}{2}\vL_k)Z_k(x)\Omega(x+\frac{1}{2}\vL_k),
$$
where $\Omega(x)\in U(N)$ is the gauge group element.

\item
In order to preserve $U(N)$ gauge invariance we have to modify
the definition of the commutators in (\ref{LagSYM}) to:
$$
\com{Z_k}{Z_l}_{(x)} \rightarrow
Z_k(x-\frac{1}{2}\vL_l)
Z_l(x+\frac{1}{2}\vL_k)
-Z_l(x-\frac{1}{2}\vL_k)
Z_k(x+\frac{1}{2}\vL_l).
$$

\item
We also need to modify the interactions of the fermions with the
scalars so as to be gauge invariant. This can be done by assigning to
the fermions their own dipole-vectors.  To find the appropriate
assignment we need to correlate the dipole-vector of the various
fields with their $Spin(6)=SU(4)$ R-symmetry charges, as follows.  The
parameters $\vL_k$ that define the dipole theory can be combined into
a single linear map $\MapL: su(4)\rightarrow \BR^{3,1}$ from the Lie
algebra of the R-symmetry group to a spacetime 4-vector.  Using the
inner product on $su(4)$, $\MapL$ can be represented as an
$su(4)$-valued spacetime 4-vector.  In the representation $\rep{6}$ of
$su(4)$ we can take $\MapL$ to be
\be\label{vLeigen}
\MapL\stackrel{\rep{6}}{\rightarrow}
\left(\begin{array}{rrrrrr}
0 &  \vL_1 & 0 & 0 & 0 & 0 \\
-\vL_1 & 0 & 0 & 0 & 0 & 0 \\
0 & 0 & 0 &  \vL_2 & 0 & 0 \\
0 & 0 & -\vL_2 & 0 & 0 & 0 \\
0 & 0 & 0 & 0 & 0 &  \vL_3 \\
0 & 0 & 0 & 0 & -\vL_3 & 0 \\
\end{array}\right).
\ee 
Now we can define the interactions of the fermions.  We need to write
$\MapL$ in the representation $\rep{4}$ of $su(4)$ and find a basis of
this representation where $\MapL$ is diagonal.  It will then have the
following form:
$$
\MapL\stackrel{\rep{4}}{\rightarrow}
\left(\begin{array}{cccc}
 \vLf_1 & 0 & 0 & 0 \\
 0 & \vLf_2 & 0 & 0 \\
 0 & 0 & \vLf_3 & 0 \\
 0 & 0 & 0 & \vLf_4 \\
\end{array}\right),
$$
with the definitions
\bear
\vLf_1 & =& \frac{1}{2}(\vL_1+\vL_2+\vL_3), \nn\\
\vLf_2 & =& \frac{1}{2}(\vL_1-\vL_2-\vL_3), \nn\\
\vLf_3 & =& \frac{1}{2}(-\vL_1+\vL_2-\vL_3), \nn\\
\vLf_4 & =& \frac{1}{2}(-\vL_1-\vL_2+\vL_3).
\eear
The Weyl fermions $\psi_\a^a$ ($a=1\dots 4$) of (\ref{LagSYM}), which
are in the $\rep{4}$ of $su(4)$, should be assigned the dipole vectors
$\vLf^a$ and their complex conjugate fields should be assigned
$(-\vLf^a)$. To get a gauge invariant Lagrangian we need
to replace all the commutators of a scalar and a fermion with:
$$
\com{Z_k}{\psi^a}_{(x)}\rightarrow
Z_k(x-\frac{1}{2}\vLf_a)\psi(x+\frac{1}{2}\vL_k)
-\psi(x-\frac{1}{2}\vL_k)Z_k(x+\frac{1}{2}\vLf_a).
$$
\item Since the gauge bosons have vanishing dipole vectors,
preserving any supersymmetry requires that some of the fermions have
vanishing dipole vectors \cite{Bergman:2001rw}. In particular, to
preserve $\SUSY{2}$ we may choose $\vLf_1= -\vLf_2=\vL_2=\vL_3=\vL$ and
$\vLf_3= \vLf_4=\vL_1=0$. 
\end{enumerate}
These rules can be recast as a redefinition of the product of two
fields. The modified product of any two fields $\Xi_1(x),\Xi_2(x)$
(scalar, fermionic or gauge) is defined in a way somewhat reminiscent
of noncommutative geometry \cite{Hoppe:1990tc,Fairlie:1988qd}:
\be\label{XiXi} 
(\Xi_1 * \Xi_2)_{(x)} \equiv e^{
  \frac{i}{2}\langle\MapL^\u,\hR_1\rangle\frac{\partial}{\partial
    z_\u}
  -\frac{i}{2}\langle\MapL^\u,\hR_2\rangle\frac{\partial}{\partial
    y_\u} }\left(\Xi_1(y)\Xi_2(z)\right)|_{y=z=x}, 
\ee 
where $\hR_i$ ($i=1,2$) is the ($su(4)$-valued) R-symmetry charge
operator acting on $\Xi_i$ and $\langle\cdot,\cdot\rangle$ is the
Killing form on $su(4)$ (see \cite{Bergman:2001rw} for more details).

Special cases of dipole theories have been discussed in
\cite{Cheung:1998te,Banks:1999tr} and various aspects of the theories
have been explored in \cite{Motl:2001dj}-\cite{Huang}.

\subsection*{Lightlike dipole-vectors}
Define the linear vector space $W\subset \BR^{3,1}$ to be the image of
the map $\MapL: su(4)\rightarrow \BR^{3,1}$ defined in (\ref{vLeigen}).
In terms of the fundamental dipole vectors
that were introduced in (\ref{vLeigen}):
$$
W = {\mbox{Span}}\{\vL_1, \vL_2, \vL_3\}.$$
We will define the dipole theory to be {\it lightlike} if $W$ is
1-dimensional and null, i.e.
$$
\vL_i\cdot\vL_j = 0,\qquad i,j=1,2,3.
$$
As we shall see in section \ref{secPropS}, lightlike dipole theories
are easier to analyze than the generic dipole theories.  This is
similar to Yang-Mills theory on a noncommutative space that simplifies
when the noncommutativity parameter is lightlike
\cite{Aharony:2000gz}.  Lightlike deformation parameters have also
been used in the context of the noncommutative $(2,0)$-theory
\cite{Aharony:1997an}-\cite{Berkooz:1998st}.

\section{Lightlike dipole theories and NSNS plane-wave backgrounds}
\label{secLLppNS}
In this section we will show that the low energy effective actions
describing appropriately oriented
D3-branes in a plane-wave background with a strong
lightlike NSNS 3-form flux are lightlike dipole theories. 
The orientation of the D3-branes must be such that, in the notation
of (\ref{ppNSi})-(\ref{ppNSiii}), the $+,-$ directions are longitudinal
and the $x^i$ ($i=6\dots 9$) directions are transverse.

\subsection{Geometric engineering of dipole-theories}
To obtain a lightlike dipole theory we consider a background in which
probe D3 branes have a small timelike dipole vector and then
we perform a large boost.
For simplicity, assume that all the dipole vectors which
are encoded in $\MapL$ are in the $x^1$ direction.  In this case
$\MapL$ reduces to a single element in the Lie algebra $su(4)$ which,
in the representation $\rep{6}$,
we can write as a $6\times 6$ antisymmetric matrix $2\pi\a'\Qh$.

As was shown in \cite{Bergman:2001rw}, a $U(N)$ dipole theory with
dipole vectors along $x^1$ described by $2\pi\a'\Qh$
arises as the low-energy
effective action of $N$ D3-brane probes in the string theory background, 
\bear 
ds^{2} &=& dt^{2} - \frac{1}{1 + \xvt\Qh^\top\Qh\xv}(dx^1)^2 - (dx^{2})^{2} -
(dx^{3})^{2} -d\xvt d\xv +\frac{(d\xvt\Qh\xv)^{2}}{1 + \xvt
  \Qh^\top\Qh\xv}
\nn\\
B &=& \frac{1}{2} \frac{d\xvt \Qh\xv}{1 + \xvt \Qh^\top \Qh\xv}\wdg
dx^1, \qquad e^{2(\phi-\phi_{0})} = \frac{1}{1 +\xvt \Qh^\top \Qh
  \xv}, 
\nn 
\eear 
where $\xv = (x^4, \ldots, x^9)$. We can obtain a theory with a
lightlike dipole vector by infinitely boosting this background along
$x^1$. As the dipole vector prior to the boost has a magnitude set by
$2\pi \a' \Qh$, we must simultaneously scale $\Qh \rightarrow 0$ to
obtain a lightlike dipole vector which has finite components in this
limit. Thus, let
$$
x^1 = \gamma({x^1}'+v t'),\quad
t = \gamma(t'+v {x^1}'),\qquad
\gamma \equiv\frac{1}{\sqrt{1-v^2}}
$$
and take $v\rightarrow 1$ while keeping
$$
\gamma\Qh \equiv Q = {\mbox{finite}}.
$$
Defining $x^\pm \equiv t'\pm {x^1}'$ we find the background
\bear
ds^{2} &=& dx^+ dx^- + (\xvt Q^\top Q\xv)(dx^+)^2 -
(dx^{2})^{2} - (dx^{3})^{2} -d\xvt d\xv
\nn\\
B &=& \frac{1}{2} d\xvt Q \xv \wdg dx^+,
\qquad
e^{\phi} = g_s.
\label{Qbckg}
\eear
In order to preserve $\SUSY{2}$ supersymmetry, we take
\be \label{L}
2\pi \a' Q =
\left(\begin{array}{cccccc}
0 &  0 & 0 & 0 & 0 & 0 \\
0 & 0 & 0 & 0 & 0 & 0 \\
0 & 0 & 0 &  L^- & 0 & 0 \\
0 & 0 & -L^- & 0 & 0 & 0 \\
0 & 0 & 0 & 0 & 0 &  L^- \\
0 & 0 & 0 & 0 & -L^- & 0 \\
\end{array}\right).
\ee 
Concretely, we note that the dipole vectors for the fields in this
background are of the form $\vL=\pm (L^-, -L^-,0,0)$. If we define
\be\label{mudef}
\mu \equiv \frac{L^-}{2\pi \a'}, 
\ee 
we see that this background (\ref{Qbckg})
is exactly the NSNS plane-wave of equation
(\ref{ppNSi})-(\ref{ppNSiii}),
\bear
ds^2 & =& dx^{+} dx^{-} + \mu x^i x^i (dx^+)^2 - dx^a dx^a - dx^i dx^i,\nn\\
H^{1} & =& -\mu dx^{+}\wdg (dx^6\wdg dx^7 + dx^8 \wdg dx^9), \nn\\
e^{\phi} & =& g_s,
\label{LLNSbg}
\eear
where again $a=2,\ldots,5$ and $i=6,\dots,9$.

Note that $L^-$, the characteristic length scale of nonlocality,
can be made arbitrarily big by a coordinate transformation that
rescales $x^{+}$.  It is therefore obvious that the excited open
string states decouple from the low energy lightlike dipole theory.
Furthermore, since the lightlike dipole theory is a limit of a dipole
theory with spacelike dipole vectors and since the latter can be
constructed as a certain limit of compactified noncommutative
$\SUSY{4}$ Super Yang-Mills theory \cite{Bergman:2000cw}, it follows
that the lightlike dipole theory is unitary.

\subsection{Lightcone string theory in the NSNS background}
Using the exact results of \cite{Berenstein:2002jq, Russo:2002rq}
(extended by \cite{Michishita:2002jp} to the open string case) for
string theory in the NSNS plane-wave background (\ref{LLNSbg}), we
will show directly that the open string interactions are modified by
the phases one would expect for a lightlike dipole deformation.

In order to facilitate future comparisons to the RR case, we consider
the plane wave background in the GS formalism. First, we define the
complex worldsheet scalar fields
$$
\wsZ_1\equiv \wsX_{6} +i\wsX_{7},\qquad
\wsZ_2\equiv \wsX_{8} +i\wsX_{9}.
$$
In order to simplify the analysis of the interactions in lightcone
gauge, it is conventional to fix $X^+ = p^{+}\tau$ and additionally
require that the string length be $\ell =2\pi \a' p^+$. 
Just as in \cite{Berenstein:2002jq, Russo:2002rq}, we will find it
useful to split our fermions into positive and negative
chirality fermions with respect to $\Gamma^{6789}$. We use $S$ to
denote the positive chirality fermions. As the negative chirality
fermions and the scalars $X^a$, $a=2,\ldots,5$ remain free and
massless, we will ignore them. The resulting action in
lightcone gauge is then given by,
\be
S = \frac{1}{2\pi \a'} \int\!\! d \t \!\int_0^{2 \pi \a' p^{+}}
\!\!\! d\sigma\,
\left\lbrack  
 \frac{1}{2}\sum_{k=1}^2
   \left(|\dot{\wsZ}_k|^2 -|\wsZ_k' + i \mu \wsZ_k|^2\right)
+i\overline{S}\left(\s^0\partial_0 
     +\s^1 ( \partial_1 - \mu \Gamma^{67}) \right) S
\right\rbrack
\label{ActPPNS}
\ee
There exists a field redefinition that, locally in $\sigma$,
transforms this action into that of a free string.
This transformation is \cite{Russo:2002rq, Michishita:2002jp}
\be\label{trPPFR}
\twsZ_k(\sigma)\equiv e^{i \mu \s}\wsZ_k(\sigma),
\qquad
\twS(\sigma)\equiv e^{-\mu \Gamma^{67}\s}S(\sigma).
\ee
In terms of the new fields the action is simply
\be
S = \frac{1}{2\pi \a'} \int d \t \int_0^{2 \pi \a' p^{+}}d\sigma\,
\left\lbrack  
 \frac{1}{2}\sum_{k=1}^2
   \left(|\dot{\twsZ}_k|^2 -|\twsZ_k'|^2\right)
 + i\overline{\twS}\left(\s^0\partial_0 
     + \s^1 \partial_1 \right) \twS
\right\rbrack
\label{ActFR}
\ee
Note that the transformation (\ref{trPPFR}) can change the boundary
conditions of various fields. For closed strings, the transformed
fields no longer satisfy periodic boundary conditions and the
closed string spectrum in the plane-wave background differs from that
of the free string. However, the spectrum of open strings with
Dirichlet boundary conditions is unaltered. Instead, the interactions
are modified in an interesting way as we will discuss presently.

\subsection{Lightlike dipole-theories on D-branes in a plane-wave background}

Consider a D1-brane that is extended in the $x^{+}$, $x^{-}$
directions. The extension of the discussion to D3-branes is
straightforward.  The open string excitations are described in
lightcone gauge by the action (\ref{ActPPNS}) with the boundary
conditions
$$
\wsZ_k(0)=\wsZ_k(2\pi\a' p^{+}) = 0,
\qquad
0 = S_L(0) - S_R(0) = 
S_L(2\pi \a'p^{+}) - S_R(2\pi \a' p^{+}).
$$
As the transformation (\ref{trPPFR}) does not affect these boundary
conditions, the spectrum of Dirichlet-Dirichlet open strings
ending on a D1-brane in this NSNS plane-wave background is the same as
the flat space spectrum. The interactions, however, receive extra
phases that precisely reproduce the interactions described in
section \ref{secDT}.
\FIGURE[ht]{
\begin{picture}(350,120)
\thinlines
\put(10,10){\vector(1,0){50}} 
\put(10,10){\vector(0,1){70}} 
\put(8,85){$\sigma$}
\put(65,10){$\tau$}

\thicklines
\put(50,20){\line(1,0){250}} %
\put(50,100){\line(1,0){250}} %
\put(50,40){\line(1,0){60}} %
\put(50,70){\line(1,0){80}} %
\put(300,60){\line(-1,0){120}} %

\qbezier(50,20)(45,25)(50,30)\qbezier(50,30)(55,35)(50,40)

\qbezier(50,40)(45,45)(50,50)\qbezier(50,50)(55,55)(50,60)
\qbezier(50,60)(45,65)(50,70)

\qbezier(50,70)(55,75)(50,80)\qbezier(50,80)(45,85)(50,90)
\qbezier(50,90)(55,95)(50,100)

\qbezier(300,20)(295,30)(300,40)\qbezier(300,40)(305,50)(300,60)

\qbezier(300,60)(295,70)(300,80)\qbezier(300,80)(305,90)(300,100)

\put(55,87){$2\pi \a' p^{+,(in)}_3$} %
\put(55,57){$2\pi \a' p^{+,(in)}_2$} %
\put(55,27){$2\pi \a' p^{+,(in)}_1$} %
\put(235,80){$2\pi \a' p^{+,(out)}_2$} %
\put(235,40){$2\pi \a' p^{+,(out)}_1$} %
\put(310,100){$2\pi \a' p^{+}$} %

\put(20,83){$V^{(in)}_3$} %
\put(20,53){$V^{(in)}_2$} %
\put(20,27){$V^{(in)}_1$} %
\put(305,80){$V^{(out)}_2$} %
\put(305,40){$V^{(out)}_1$} %
\end{picture} \label{Fig1}
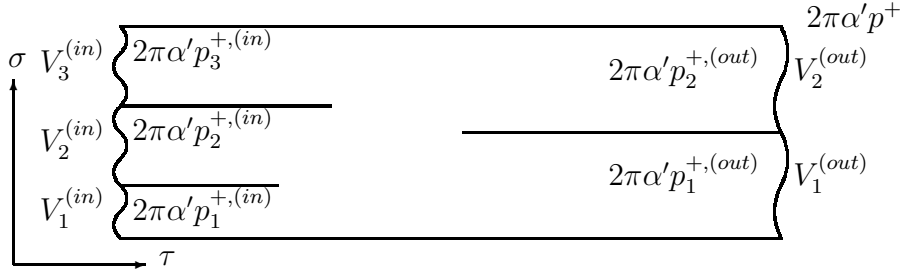
\caption{Scattering amplitude in the lightcone formalism.}
}
Consider, for example, a tree level diagram that describes the
scattering of open string states with vertex operators
$V^{(in)}_1,\cdots, V^{(in)}_{n_i}$ into open string states with
vertex operators $V^{(out)}_1,\cdots,V^{(out)}_{n_f}$ (see Figure
\ref{Fig1}).  When written in terms of $\twsZ_k$ and $\twS$, these
vertex operators should have the same form as the usual free
Dirichlet-Dirichlet open string vertex operators. In fact, one might
naively guess that as $\twsZ_k = e^{i \mu \s } \wsZ_k$
the relation should be
\be\label{VVfr}
V^{(in)}_j(\wsZ_k(\sigma),\dots) =
\tV^{(in)}_j(
e^{-i\mu \sigma}\wsZ_k(\sigma),\dots),
\ee
where $\tV^{(in)}_j$ is the free string vertex operator that
corresponds to the free string state with the same labels. This, of
course, would give us the same amplitudes as in the free string case.
However, note that if we let $p^{+,(in)}_j$ be the lightcone momentum
of the $j^{th}$ incoming string state, the parameter $\sigma$ for that
state is in the range
$$
2\pi \a' \sum_{k=1}^{j-1} p^{+,(in)}_k\le
\sigma\le 2\pi \a' \sum_{k=1}^j p^{+,(in)}_k,
$$
which means that the prescription (\ref{VVfr}) for defining the vertex
operator contains phase factors which depend on the position of the
insertion of the operator along the string. This cannot be correct. 

We can solve this problem by replacing $\s$ with $\s' = \s -
2\pi \a' \sum_{k=1}^{j-1} p^{+,(in)}_k$ (which is the distance from
the beginning of the $j^{th}$ string) so 
\bear
0 \le & \s' & \le 2\pi \a' p^{+,(in)}_j \\
V^{(in)}_j(\wsZ_k(\sigma),\dots) & = &
\tV^{(in)}_j(
e^{-i\mu \s'} \wsZ_k(\sigma),\dots). \label{VVpp}
\eear
This modification leads to overall phase shifts in the vertex
operators as compared to the theory in flat space. To calculate them,
we just need to know the $\wsZ_k$ and $S$ dependence of the vertex
operators. More formally, on the worldsheet there is a global $U(1)$
symmetry which acts on the $\wsZ_k$ by $\wsZ_k \rightarrow
e^{i\theta}\wsZ_k$ (and analogously on the fermions, which we neglect for
simplicity). A general vertex operator will transform under this
$U(1)$ as $V^{(j)} \rightarrow e^{iq^{(j)} \theta} V^{(j)}$.
Noting that $\mu = \frac{L^-}{2\pi\a'}$, it is easy to see that the
definitions (\ref{VVpp}) and (\ref{VVfr}) differ by the phase,
$$
\exp\left\{ i \sum_{l=1}^{j-1} q^{(j)} L^- p^{+,(in)}_l
\right\}.
$$
If we let $p^{+,(out)}_r$ be the lightcone momentum of the $r^{th}$
outgoing string state, momentum conservation requires $p^{+} =
\sum_{j=1}^{n_i} p^{+,(in)}_j =\sum_{r=1}^{n_f} p^{+,(in)}_r$ and we
see that the overall phase for the entire amplitude is
$$
\exp\left\{
i \sum_{j=1}^{n_i}\sum_{l=1}^{j-1} q^{j,(in)} L^- p^{+,(in)}_l
-i\sum_{r=1}^{n_f}\sum_{s=1}^{r-1} q^{r,(out)} L^- p^{+,(out)}_s
\right\}.
$$
It is not hard to see that this is exactly the same phase as the
one we get by Fourier expanding the Super Yang-Mills action
of a D-brane and replacing every product with the modified
$*$-product (\ref{XiXi}).

\section{Proposal for the S-dual theory}
\label{secPropS}
In this section we will present our proposal for the S-dual of
the lightlike dipole theories. We will begin with an analysis
of a dipole theory with a $U(1)$ gauge group and a single fermion
(known as dipole QED \cite{Dasgupta:2001zu, Sadooghi:2002ph}) and
then proceed to present our conjecture about a dipole theory with
an $SU(N)$ or $U(N)$ gauge group.

The field contents of $U(1)$ dipole QED 
(without any supersymmetry) is:
\vskip 0.5cm
\begin{tabular}{ll}
$A_\u$ & the $U(1)$ gauge field, \\
$\psi$ & a Dirac fermion with dipole vector $\vL$. \\
\end{tabular}
\vskip 0.5cm
\noindent The Lagrangian is
\bear
\Lag &=& \frac{1}{4g^2} F_{\u\v}F^{\u\v}
+\frac{1}{2g^2} \bpsi\gamma^\u D_\u\psi,\qquad
D_\u\psi\equiv \px{\u}\psi -i \lbrack 
A_\u(x+\frac{1}{2}\vL)-A_\u(x-\frac{1}{2}\vL)
\rbrack\psi.
\nn\\
\label{QEDLag}
\eear
Here $\vL$ is the constant dipole-vector and we assume that it
is spacelike or null. 

As shown in \cite{Dasgupta:2001zu, Sadooghi:2002ph}, the Feynman rules
of this theory are identical to those of ordinary QED, with the
following modification of the interaction vertex,
\begin{equation}
ig\g^\mu \rightarrow ig\g^\mu
  \times 2i \sin{\frac{p\cdot L}{2}},
\end{equation}
where $p$ is the outgoing momentum of the photon. In particular,
this means that the photon self-energy at one-loop just gets an extra
factor of 
\be
2i\sin{\frac{p\cdot L}{2}}\times
 2i \sin{\frac{-p\cdot L}{2}} =
4 \sin^2 {\frac{p\cdot L}{2}}
\ee
as compared to the QED result. This suggests that the $U(1)$ theory is
IR free, just like ordinary QED. Thus, our application of S-duality in
the $U(1)$ case will be somewhat formal, and should be considered simply
as a motivation for the conjecture in the $U(N)$ case.

To find the S-dual description we will adopt the standard method of
using a Lagrange multiplier for the field strength.\footnote{A similar
  method was used in \cite{Ganor:2000my} to study S-duality for Super
  Yang-Mills theory on a noncommutative $R^{3,1}$.} We treat
$F_{\u\v}$ as an independent field subject to the Bianchi identity
$\epsilon^{\u\v\t\r}\px{\v}F_{\t\r} = 0$ which we implement with a
Lagrange multiplier.  Of course, this method requires that the gauge
field $A_\u$ does not appear explicitly in the Lagrangian. Unlike in
ordinary QED, here we can eliminate $A_\u$ by performing a
redefinition of variables \cite{Bergman:2000cw}
\be\label{npsidef}
\npsi(x) \equiv e^{-\frac{i}{2}\int_{-1}^1 \cvL{\u} A_\u
  (x+\frac{s}{2}\vL)ds} \psi(x), 
\ee 
so that $\npsi$ is a $U(1)$-neutral field. This is the analog of the
Seiberg-Witten map \cite{Seiberg:1999vs} for dipole theories. Just as
in that case, this transformation results in a theory with ordinary
gauge symmetry perturbed by an infinite number of irrelevant
interactions. In particular, since
$$
D_\u\psi(x) = e^{-\frac{i}{2}\int_{-1}^1 \cvL{\v}A_\v (x+\frac{s}{2}\vL)ds}
\left\lbrack
\px{\u}\npsi(x)
+\frac{i}{2}\npsi(x)\int_{-1}^1 \cvL{\v} F_{\u\v}(x+\frac{s}{2}\vL)ds
\right\rbrack,
$$
we can define
$$
\nDF{\u}\npsi(x)\equiv
\px{\u}\npsi(x)
+\frac{i}{2}\npsi(x)\int_{-1}^1 \cvL{\v} F_{\u\v}(x+\frac{s}{2}\vL)ds,
$$
to get the Lagrangian
\be \label{AfterSW}
\Lag_1 = \frac{1}{4g^2} F_{\u\v}F^{\u\v}
+\frac{1}{2g^2} \bnpsi\gamma^\u \nDF{\u}\npsi.
\ee
Thus, as promised, the explicit dependence on $A_\u$ has been removed.
Further, note that making the replacements $\psi \rightarrow
g \psi$, and $A_\mu \rightarrow g A_\mu$ in the above Lagrangian and
writing 
\be \label{Intfor}
\int_{-1}^1 ds F_{\u\v}(x+\frac{s}{2}\vL)=
\frac{2\sin{\frac{i}{2}L\cdot\partial}}{
            \frac{i}{2}L\cdot\partial} F_{\u\v}(x),
\ee
we see that
$$
\Lag_1 = \frac{1}{4} F_{\u\v}F^{\u\v}
+\frac{1}{2} \bnpsi\gamma^\u \left( \partial_\u +igL^\v
  \frac{\sin{\frac{i}{2}L\cdot\partial}}{
             \frac{i}{2}L\cdot\partial} F_{\u\v}(x) \right)\npsi
$$
is just a free theory perturbed by an infinite number of higher
derivative interactions with couplings of the form $gL\times L^{2n}$.

We can now easily find the S-dual theory by treating $F_{\u\v}$
as an independent variable and adding a Lagrange multiplier to the
Lagrangian (\ref{AfterSW}),
$$
\Lag_2 = \frac{1}{4g^2} F_{\u\v}F^{\u\v}
+\frac{1}{2g^2} \bnpsi\gamma^\u \nDF{\u}\npsi
+\frac{1}{8\pi} \wA_\u\epsilon^{\u\v\t\r}\px{\v}F_{\t\r}.
$$
Except for the kinetic term, $F_{\u\v}$ appears linearly in
$\Lag_2$. Thus, we can integrate it out to get,
\be
\Lag' = 
\frac{1}{4{g'}^2}
\left(\wF_{\u\v}
-\frac{1}{\pi {g'}^2}
\epsilon_{\u\v\t\r}\cvL{\t}\int_{-1}^1 \wJ^\r (x+\frac{s}{2}\vL)ds
\right)^2
+\frac{1}{2{g'}^2} \bwpsi\gamma^\u \px{\u}\wpsi.
\label{Lagdl}
\ee
where we have defined
\be
g'=\frac{4\pi}{g}, \qquad \wpsi \equiv \frac{4\pi}{g^2}\psi, \qquad
\wJ_\u \equiv i\bwpsi\gamma_\u\wpsi.
\ee
If we make a further redefinition of the fields,
\be\label{hLdef}
\hpsi \equiv \frac{1}{g'} \wpsi, \qquad \hF \equiv \frac{1}{g'} \wF,
\qquad \hL^\t \equiv \frac{1}{{g'}^2} L^\t,
\ee
we see that (\ref{Lagdl}) can be rewritten as
\be
\Lag' = 
\frac{1}{4}
\left(\hF_{\u\v}
-\frac{g'\hL^\t}{\pi}
\epsilon_{\u\v\t\r}\int_{-1}^1 \hJ^\r (x+\frac{s}{2}{g'}^2\hL)ds
\right)^2
+\frac{1}{2} \bhpsi \gamma^\u \px{\u}\hpsi.
\label{Lagdual}
\ee
If we add minimally coupled scalars to the QED Lagrangian
(\ref{QEDLag}), with the same dipole vector $\vL$,
 the expression of the S-dual $\Lag'$ becomes 
more complicated because the interactions are quadratic in the 
gauge field.
The dual Lagrangian simplifies for a lightlike dipole vector.
To see this, we will fix the QED lightcone gauge $A_{-} = 0$.
In this gauge the redefinition (\ref{npsidef}) becomes simply
$\npsi(x) \equiv \psi(x)$.
Following the same steps that led to (\ref{Lagdl}) with minimally
coupled scalars added we find that the dual Lagrangian can be
obtained from the QED Lagrangian by the substitution
\be\label{GenSubst}
g\rightarrow g',\qquad
F_{\u\v}(x)\rightarrow
F'_{\u\v}(x)\equiv F_{\u\v}(x)
-\frac{\hL^\t}{\pi}
\epsilon_{\u\v\t\r}\int_{-1}^1 \hJ^\r (x+\frac{s}{2}{g'}^2\hL)ds,
\ee
where $\hJ^\r$ is the $U(1)$ current including the contribution 
of the scalars.

We see that the S-dual theory actually
looks local order by order in $g'$, and only appears non-local if we
sum all orders in $g'$. In particular, the scale of non-locality in
this description is ${g'}^2\hL$.
We can gain a clue as to the origin of
the non-locality by rewriting (\ref{GenSubst}) using (\ref{Intfor})
\be
F_{\u\v}'(x) = 
F_{\u\v}(x)
-\frac{\hL^\t}{\pi}\epsilon_{\u\v\t\r} 
   \frac{2\sin{\frac{i}{2}{g'}^2\hL \cdot \partial}}{
               \frac{i}{2}{g'}^2\hL \cdot \partial} \hJ^\r (x).
\ee
Notice that only even powers of ${g'}^2$ enter in 
the Taylor series expansion of 
$\frac{\sin{\frac{i}{2}{g'}^2\hL \cdot \partial}}{
\frac{i}{2}{g'}^2\hL \cdot \partial}.$
It would be interesting to understand
this behavior directly by studying string interactions in the
S-dual RR plane wave background (\ref{ppRi})-(\ref{ppRiii}). 
Note that when the NSNS background (\ref{ppNSi})-(\ref{ppNSiii})
is transformed into the RR background (\ref{ppRi})-(\ref{ppRiii})
using S-duality, the Regge slope $\talp$ of (\ref{ppRi})-(\ref{ppRiii})
is given in terms of the Regge slope $\a'$ of
(\ref{ppNSi})-(\ref{ppNSiii}) by $\talp=g_s\a'$.
Using (\ref{mudef}) and the definition of $\hL$ in (\ref{hLdef})
we see that $\hL=2\pi \talp\mu$ and so is finite in the RR background.

In order to extend the discussion to $N$ D3-brane probes we need to know
the S-dual description of the dipole theory that is obtained as
a deformation of $\SUSY{4}$ Super Yang-Mills theory with gauge group $U(N)$.
Since the gauge fields that correspond to the $U(1)$ center are IR free
we can ignore them and consider only the $SU(N)$ dipole theory.
It is natural to conjecture that the dual of the lightlike $SU(N)$
dipole theory is given by a prescription similar to (\ref{GenSubst})
\bear
F_{\u\v}(x) &\rightarrow&
F_{\u\v}(x)
-\frac{1}{\pi}\epsilon_{\u\v\t\r}\int_{-1}^1 ds
\left\{
\sum_{k=1}^3\hL_k^\t\hJ_k^\r (x+\frac{s}{2}{g'}^2\hL_k)
+\sum_{a=1}^4\hLf_a^\t\hJ_a^\r (x+\frac{s}{2}{g'}^2\hLf_a)
\right\},
\nn\\
\hJ_k^\u &\equiv& \frac{1}{2}(i\bZ_k D^\u Z_k -i D^\u\bZ_k Z_k
-iZ_k D^\u\bZ_k +i D^\u Z_k\bZ_k),
\nn\\
\hJ_a^\u &\equiv& 
 \frac{1}{2}{\sigma^{\u\a}}_\da\psi^a_\a\bpsi^\da_a
-\frac{1}{2}{\sigma^{\u\a}}_\da\bpsi^\da_a\psi^\a_\a
\label{SUSubst}
\eear
Here we used the notation of section \ref{secDT} and we defined
the rescaled dipole vectors of the bosons and fermions 
similarly to (\ref{hLdef}),
$$
\hL_k^\t \equiv \frac{1}{{g'}^2} L_k^\t,\qquad
\hLf_a^\t \equiv \frac{1}{{g'}^2} \lambda_a^\t.
$$
The Lie algebra valued $\hJ_k^\u$ and $\hJ_a^\u$
are the individual contributions
of the scalars and fermions to the $su(N)$ current.
In (\ref{SUSubst}) each contribution to the
current enters with a coefficient that
is proportional to the dipole vector of the corresponding field.
We assume that the dipole vectors $\hL_k$ and $\hLf_a$ are all
lightlike and pointing in the same direction.
We also assume that (\ref{SUSubst}) is written in the gauge $A_{-}=0$.
In this case all the residual
gauge transformations are independent of $x^{-}$ and 
(\ref{SUSubst}) is gauge invariant.

We do not know what should be the modification to the potential of the scalar fields
and the Yukawa coupling of the scalars and fermions.
It is possible that those interactions are still given by the $*$-product
modification (\ref{XiXi}) with the unrescaled dipole vectors $L_k$ and $\lambda_a$
(which are now of order ${g'}^2$).
Although the scalars and fermions are not expected to transform as dipoles under
the ``dual'' gauge fields (since they are electric but not magnetic dipoles),
in the $A_{-}=0$ gauge the interactions are gauge invariant even after the modification
(\ref{XiXi}).

\section{Conclusion and discussion}\label{secDisc}
In this paper we have argued that particular D-brane probes of
plane-wave backgrounds are described by nonlocal field theories.
In the case of an NSNS background we have identified the field theory
as a lightlike dipole theory and we have verified the statement by
an explicit lightcone string computation.
In the case of an RR 3-form field strength background we have provided
an indirect argument, using S-duality, for the nonlocality of the 
effective theory on D3-brane probes. 
This is a more complicated theory and we have conjectured the form
of its Lagrangian in (\ref{GenSubst}). 
The nonlocality scale is proportional to $g_s$ and 
it is obvious from (\ref{GenSubst})
that one has to sum up contributions from
all orders of string perturbation theory in order to exhibit 
the nonlocal nature of the interactions.
It would be interesting to verify this directly from the solvable
plane-wave string theory.\footnote{Work in progress.}
Note that, since the nonlocal interactions are in the lightlike
direction, we can make the characteristic scale arbitrarily big
by a coordinate transformation that rescales $x^{\pm}$. 
The excited string states can therefore decouple safely and, as
the field theoretic S-duality suggests, the effective nonlocal
field theory can be unitary.

It is interesting to extend these ideas to pp-wave backgrounds with
other RR fluxes. For that purpose we adopt the following somewhat
heuristic point of view.
The dipole theories that we have described in section \ref{secDT}
have a correlation between R-symmetry charge and electric flux.
In the D3-brane language, every state with $Spin(6)$ transverse
angular momentum also behaves as a fundamental string of finite
extent. The length of the string is proportional to its angular
momentum and the proportionality constants are the dipole
vectors $\vL_k$.
In the S-dual nonlocal theories that describe D3-brane probes
in pp-waves with a 3-form RR flux 
every state with $Spin(6)$ transverse
angular momentum also behaves as a D1-brane of finite
extent.
We can extend this line of thought to other RR-backgrounds.
For example, in a background with a 5-form RR field strength
$F_{+1234}=F_{+5678}$ 
(where we use lightcone coordinates $+,-,1\dots 8$ as in
\cite{Berenstein:2002jq}) 
and a D5-brane in directions $+,-,1235$ we should find that
open string states 
attached to the D5-brane that have, say, angular
momentum in the $78$ plane also behave as a D3-brane
that is spread in directions $+,-,56$ and has a finite volume
that is proportional to the angular momentum.
This statement is, admittedly, obscure and 
it would be interesting to elucidate such a theory further.

Another possible application of the ideas presented in this paper
is to M(atrix)-theory \cite{Banks:1996vh}.
The M(atrix)-theory Hamiltonian for M-theory is the 0+1D supersymmetric
Yang-Mills quantum mechanics \cite{Baake:1984ie}-\cite{deWit:1988ig}.
The standard derivation of weakly coupled type-IIA string theory
from M(atrix)-theory \cite{Motl:1997th}-\cite{Dijkgraaf:1997vv}
requires understanding of the strong coupling limit of 1+1D 
$\SUSY{8}$ Super Yang-Mills theory.
Dipole theories naturally appear as M(atrix)
models of Melvin spaces \cite{Motl:2001dj} (see also 
\cite{Cheung:1998te,Banks:1999tr}). The relevant M(atrix) models are
dipole theories that are deformations of 1+1D $\SUSY{8}$ Super Yang-Mills 
theory. Therefore understanding the strong coupling limit of dipole theories
could prove beneficial for deriving a weakly coupled string theory descriptions
of Melvin spaces. (A string theory description for Melvin backgrounds
has been given in \cite{Russo:1998xv} but it has a dilaton that is not bounded.)
Perhaps nonlocal worldsheet theories will
play a role in such a description.
(See \cite{Aharony:2001pa}-\cite{Adams:2002ft}  for other ideas regarding nonlocal
worldsheet theories.)

We would also like to mention another new kind of nonlocal theory
that appears on D3-brane probes in certain backgrounds with strong
NSNS flux \cite{Hashimoto:2002nr}-\cite{Dolan:2002px}.
It is a very intriguing nonlocal field theory 
that is not translationally invariant and is described as 
a gauge theory on a noncommutative space with a varying 
noncommutativity parameter.

\acknowledgments
It is a pleasure to thank Allan Adams, Aaron Bergman, Keshav Dasgupta, 
Jaume Gomis, Shahin Sheikh-Jabbari, Anatoly Konechny, 
Eliezer Rabinovici and Mukund Rangamani for helpful discussions.
The work of OJG was supported in part by the Director, Office of Science,
Office of High Energy and Nuclear Physics, of the U.S. Department of
Energy under Contract DE-AC03-76SF00098, and in part by
the NSF under grant PHY-0098840.




\begin{thebibliography}{1}

\bibitem{Bergman:2000cw}
A.~Bergman and O.~J.~Ganor,
{``Dipoles, twists and noncommutative gauge theory,''}
JHEP {\bf 0010}, 018 (2000)
[arXiv:hep-th/0008030].

\bibitem{Dasgupta:2000ry}
K.~Dasgupta, O.~J.~Ganor and G.~Rajesh,
{``Vector deformations of N = 4 super-Yang-Mills theory,
pinned branes,  and arched strings,''}
JHEP {\bf 0104}, 034 (2001)
[arXiv:hep-th/00101095].

\bibitem{Bergman:2001rw}
A.~Bergman, K.~Dasgupta, O.~J.~Ganor, J.~L.~Karczmarek and G.~Rajesh,
{``Nonlocal field theories and their gravity duals,''}
Phys.\ Rev.\ D {\bf 65}, 066005 (2002)
[arXiv:hep-th/0103090].

\bibitem{Maldacena:1997re}
J.~M.~Maldacena,
{``The large $N$ limit of superconformal field theories and supergravity,''}
Adv.\ Theor.\ Math.\ Phys.\  {\bf 2}, 231 (1998)
[Int.\ J.\ Theor.\ Phys.\  {\bf 38}, 1113 (1999)]
[arXiv:hep-th/9711200].

\bibitem{Berenstein:2002jq}
D.~Berenstein, J.~M.~Maldacena and H.~Nastase,
{``Strings in flat space and pp waves from $N=4$ super Yang Mills,''}
JHEP {\bf 0204}, 013 (2002)
[arXiv:hep-th/0202021].

\bibitem{Metsaev:2001bj}
R.~R.~Metsaev,
{``Type IIB Green-Schwarz superstring in plane wave Ramond-Ramond  background,''}
Nucl.\ Phys.\ B {\bf 625}, 70 (2002)
[arXiv:hep-th/0112044].

\bibitem{Metsaev:2002re}
R.~R.~Metsaev and A.~A.~Tseytlin,
{``Exactly solvable model of superstring in plane wave Ramond-Ramond  background,''}
Phys.\ Rev.\ D {\bf 65}, 126004 (2002)
[arXiv:hep-th/0202109].

\bibitem{Russo:2002rq}
J.~G.~Russo and A.~A.~Tseytlin,
{``On solvable models of type IIB superstring in NS-NS and R-R plane wave  backgrounds,''}
JHEP {\bf 0204}, 021 (2002)
[arXiv:hep-th/0202179].

\bibitem{Lunin:2002fw}
O.~Lunin and S.~D.~Mathur,
{``Rotating deformations of $AdS_3\times S^3$,
the orbifold CFT and strings in  the pp-wave limit,''}
Nucl.\ Phys.\ B {\bf 642}, 91 (2002)
[arXiv:hep-th/0206107].

\bibitem{Gomis:2002qi}
J.~Gomis, L.~Motl and A.~Strominger,
{``pp-wave / CFT$_2$ duality,''}
arXiv:hep-th/0206166.

\bibitem{Billo:2002ff}
M.~Billo and I.~Pesando,
{``Boundary states for GS superstrings in an Hpp wave background,''}
Phys.\ Lett.\ B {\bf 536}, 121 (2002)
[arXiv:hep-th/0203028].

\bibitem{Dabholkar:2002zc}
A.~Dabholkar and S.~Parvizi,
{``Dp branes in pp-wave background,''}
arXiv:hep-th/0203231.

\bibitem{Bergman:2002hv}
O.~Bergman, M.~R.~Gaberdiel and M.~B.~Green,
{``D-brane interactions in type IIB plane-wave background,''}
arXiv:hep-th/0205183.

\bibitem{Chu:2002in}
C.~S.~Chu and P.~M.~Ho,
{``Noncommutative D-brane and open string in
pp-wave background with  B-field,''}
Nucl.\ Phys.\ B {\bf 636}, 141 (2002)
[arXiv:hep-th/0203186].

\bibitem{Lee:2002cu}
P.~Lee and J.~W.~Park,
{``Open strings in PP-wave background
from defect conformal field theory,''}
arXiv:hep-th/0203257.

\bibitem{Berenstein:2002zw}
D.~Berenstein, E.~Gava, J.~M.~Maldacena, K.~S.~Narain and H.~Nastase,
{``Open strings on plane waves and their Yang-Mills duals,''}
arXiv:hep-th/0203249.

\bibitem{Balasubramanian:2002sa}
V.~Balasubramanian, M.~X.~Huang, T.~S.~Levi and A.~Naqvi,
{``Open strings from $N = 4$ super Yang-Mills,''}
arXiv:hep-th/0204196.

\bibitem{Kumar:2002ps}
A.~Kumar, R.~R.~Nayak and Sanjay,
{``D-brane solutions in pp-wave background,''}
arXiv:hep-th/0204025.

\bibitem{Alishahiha:2002rw}
M.~Alishahiha and A.~Kumar,
{``D-brane solutions from new isometries of pp-waves,''}
Phys.\ Lett.\ B {\bf 542}, 130 (2002)
[arXiv:hep-th/0205134].

\bibitem{Skenderis:2002vf}
K.~Skenderis and M.~Taylor,
{``Branes in AdS and pp-wave spacetimes,''}
JHEP {\bf 0206}, 025 (2002)
[arXiv:hep-th/0204054].

\bibitem{Michishita:2002jp}
Y.~Michishita,
{``D-branes in NSNS and RR pp-wave backgrounds and S-duality,''}
arXiv:hep-th/0206131.

\bibitem{Sugiyama:2002tf}
K.~Sugiyama and K.~Yoshida,
{``Type IIA string and matrix string on pp-wave,''}
arXiv:hep-th/0208029.

\bibitem{Hyun:2002fk}
S.~j.~Hyun and J.~H.~Park,
{``5D action for longitudinal five branes on a pp-wave,''}
arXiv:hep-th/0209219.

\bibitem{Hoppe:1990tc}
J.~Hoppe,
{``Membranes And Integrable Systems,''}
Phys.\ Lett.\ B {\bf 250}, 44 (1990).

\bibitem{Fairlie:1988qd}
D.~B.~Fairlie, P.~Fletcher and C.~K.~Zachos,
{``Trigonometric Structure Constants For New Infinite Algebras,''}
Phys.\ Lett.\ B {\bf 218}, 203 (1989).

\bibitem{Cheung:1998te}
Y.~K.~Cheung, O.~J.~Ganor and M.~Krogh,
{``On the twisted (2,0) and little-string theories,''}
Nucl.\ Phys.\ B {\bf 536}, 175 (1998)
[arXiv:hep-th/9805045].

\bibitem{Banks:1999tr}
T.~Banks and L.~Motl,
{``A nonsupersymmetric matrix orbifold,''}
JHEP {\bf 0003}, 027 (2000)
[arXiv:hep-th/9910164].

\bibitem{Motl:2001dj}
L.~Motl,
{``Melvin matrix models,''}
arXiv:hep-th/0107002.

\bibitem{Dasgupta:2001zu}
K.~Dasgupta and M.~M.~Sheikh-Jabbari,
{``Noncommutative dipole field theories,''}
JHEP {\bf 0202}, 002 (2002)
[arXiv:hep-th/0112064].

\bibitem{Correa:2002cd}
D.~H.~Correa, G.~S.~Lozano, E.~F.~Moreno and F.~A.~Schaposnik,
{``Anomalies in noncommutative dipole field theories,''}
JHEP {\bf 0202}, 031 (2002)
[arXiv:hep-th/0202040].

\bibitem{Alishahiha:2002ex}
M.~Alishahiha and H.~Yavartanoo,
{``Supergravity description of the large
$N$ noncommutative dipole field  theories,''}
JHEP {\bf 0204}, 031 (2002)
[arXiv:hep-th/0202131].

\bibitem{Sadooghi:2002ph}
N.~Sadooghi and M.~Soroush,
{``Noncommutative dipole QED,''}
arXiv:hep-th/0206009.

\bibitem{Huang}
W.-H.~Huang,
{``Exact Wavefunction in a Noncommutative Dipole Field Theory,''}
arXiv:hep-th/0208199.

\bibitem{Aharony:2000gz}
O.~Aharony, J.~Gomis and T.~Mehen,
{``On theories with light-like noncommutativity,''}
JHEP {\bf 0009}, 023 (2000)
[arXiv:hep-th/0006236].

\bibitem{Aharony:1997an}
O.~Aharony, M.~Berkooz and N.~Seiberg,
{``Light-cone description of (2,0) 
   superconformal theories in six  dimensions,''}
Adv.\ Theor.\ Math.\ Phys.\  {\bf 2}, 119 (1998)
[arXiv:hep-th/9712117].

\bibitem{Nekrasov:1998ss}
N.~Nekrasov and A.~Schwarz,
{``Instantons on noncommutative $R^4$
and (2,0) superconformal six  dimensional theory,''}
Commun.\ Math.\ Phys.\  {\bf 198}, 689 (1998)
[arXiv:hep-th/9802068].

\bibitem{Berkooz:1998st}
M.~Berkooz,
{``Non-local field theories and the non-commutative torus,''}
Phys.\ Lett.\ B {\bf 430}, 237 (1998)
[arXiv:hep-th/9802069].



\bibitem{Ganor:2000my}
O.~J.~Ganor, G.~Rajesh and S.~Sethi,
{``Duality and non-commutative gauge theory,''}
Phys.\ Rev.\ D {\bf 62}, 125008 (2000)
[arXiv:hep-th/0005046].

\bibitem{Seiberg:1999vs}
N.~Seiberg and E.~Witten,
{``String theory and noncommutative geometry,''}
JHEP {\bf 9909}, 032 (1999)
[arXiv:hep-th/9908142].

\bibitem{Banks:1996vh}
T.~Banks, W.~Fischler, S.~H.~Shenker and L.~Susskind,
{``M theory as a matrix model: A conjecture,''}
Phys.\ Rev.\ D {\bf 55}, 5112 (1997)
[arXiv:hep-th/9610043].

\bibitem{Baake:1984ie}
M.~Baake, M.~Reinicke and V.~Rittenberg,
{``Fierz Identities For Real Clifford Algebras
And The Number Of Supercharges,''}
J.\ Math.\ Phys.\  {\bf 26}, 1070 (1985).

\bibitem{Flume:1984mn}
R.~Flume,
{``On Quantum Mechanics With Extended Supersymmetry
And Nonabelian Gauge Constraints,''}
Annals Phys.\  {\bf 164}, 189 (1985).

\bibitem{Claudson:1984th}
M.~Claudson and M.~B.~Halpern,
{``Supersymmetric Ground State Wave Functions,''}
Nucl.\ Phys.\ B {\bf 250}, 689 (1985).

\bibitem{deWit:1988ig}
B.~de Wit, J.~Hoppe and H.~Nicolai,
{``On The Quantum Mechanics Of Supermembranes,''}
Nucl.\ Phys.\ B {\bf 305}, 545 (1988).

\bibitem{Motl:1997th}
L.~Motl,
{``Proposals on nonperturbative superstring interactions,''}
arXiv:hep-th/9701025.

\bibitem{Banks:1996my}
T.~Banks and N.~Seiberg,
{``Strings from matrices,''}
Nucl.\ Phys.\ B {\bf 497}, 41 (1997)
[arXiv:hep-th/9702187].

\bibitem{Dijkgraaf:1997vv}
R.~Dijkgraaf, E.~Verlinde and H.~Verlinde,
{``Matrix string theory,''}
Nucl.\ Phys.\ B {\bf 500}, 43 (1997)
[arXiv:hep-th/9703030].

\bibitem{Russo:1998xv}
J.~G.~Russo and A.~A.~Tseytlin,
{``Green-Schwarz superstring action in
a curved magnetic Ramond-Ramond  background,''}
JHEP {\bf 9804}, 014 (1998)
[arXiv:hep-th/9804076].

\bibitem{Aharony:2001pa}
O.~Aharony, M.~Berkooz and E.~Silverstein,
{``Multiple-trace operators and non-local string theories,''}
JHEP {\bf 0108}, 006 (2001)
[arXiv:hep-th/0105309].

\bibitem{Aharony:2001dp}
O.~Aharony, M.~Berkooz and E.~Silverstein,
{``Non-local string theories on $AdS_3 \times S^3$
 and stable  non-supersymmetric backgrounds,''}
Phys.\ Rev.\ D {\bf 65}, 106007 (2002)
[arXiv:hep-th/0112178].

\bibitem{Adams:2002ft}
A.~Adams, J.~McGreevy and E.~Silverstein,
{``Decapitating Tadpoles,''}
arXiv:hep-th/0209226.

\bibitem{Hashimoto:2002nr}
A.~Hashimoto and S.~Sethi,
{``Holography and String Dynamics in Time-dependent Backgrounds,''}
arXiv:hep-th/0208126.

\bibitem{Alishahiha:2002bk}
M.~Alishahiha and S.~Parvizi,
{``Branes in time-dependent backgrounds and AdS/CFT correspondence,''}
arXiv:hep-th/0208187.

\bibitem{Dolan:2002px}
L.~Dolan and C.~R.~Nappi,
{``Noncommutativity in a time-dependent background,''}
arXiv:hep-th/0210030.



































\end{thebibliography}
\end{document}